\documentclass[twocolumn]{aastex63}
\usepackage{amsmath}
\usepackage{multirow}
\usepackage{tablefootnote}

\def \nustar {{\em NuSTAR}}
\def \fluxcgs {$\mathrm{erg\,cm^{-2}\,s^{-1}}$}
\def \lumcgs {$\mathrm{erg\,s^{-1}}$}
\def \grs {GRS~1741.9–2853}
\def \caltech {{Cahill Center for Astronomy and Astrophysics, California Institute of Technology, Pasadena, CA 91125, USA}}

\newcommand{\code}{\texttt}
\newcommand{\comment}[1]{}

\shorttitle{Type-I Bursts from \grs}
\shortauthors{Pike et al.}
\graphicspath{{./}{}}
\begin{document}

\title{Photospheric Radius Expansion and a double-peaked type-I X-ray burst from \grs}

\correspondingauthor{Sean N. Pike}
\email{spike@caltech.edu}

\author[0000-0002-8403-0041]{Sean N. Pike}
\affiliation{\caltech{}}

\author{Fiona A. Harrison}
\affiliation{\caltech{}}

\author{John A. Tomsick}
\affiliation{Space Sciences Laboratory, University of California, 7 Gauss Way, Berkeley, CA 94720-7450, USA}

\author{Matteo Bachetti}
\affiliation{INAF-Osservatorio Astronomico di Cagliari, via della Scienza 5, I-09047 Selargius, Italy}

\author{Douglas J. K. Buisson}
\affiliation{Department of Physics and Astronomy, University of Southampton, Highfield, Southampton, SO17 1BJ}

\author[0000-0003-3828-2448]{Javier A. Garc\'ia}
\affiliation{\caltech{}}
\affiliation{Dr. Karl Remeis-Observatory and Erlangen Centre for Astroparticle Physics, Sternwartstr.~7, 96049 Bamberg, Germany}

\author[0000-0002-9639-4352]{Jiachen Jiang}
\affiliation{Department of Astronomy, Tsinghua University, Shuangqing Road 30, Beijing 100034, China}
\affiliation{Tsinghua Center for Astrophysics, Tsinghua University, Shuangqing Road 30, Beijing 100034, China}

\author[0000-0002-8961-939X]{R.~M.~Ludlam}\thanks{NASA Einstein Fellow}
\affiliation{\caltech{}}

\author[0000-0003-1252-4891]{Kristin K. Madsen}
\affiliation{CRESST and X-ray Astrophysics Laboratory, NASA Goddard Space Flight Center, Greenbelt, MD 20771, USA}

% \nocollaboration{2}



\submitjournal{ApJ}
\accepted{June 24, 2021}

\begin{abstract}

\noindent We present analysis of two type-I X-ray bursts observed by \nustar\ originating from the very faint transient neutron star low-mass X-ray binary \grs\ during a period of outburst in May 2020. We show that the persistent emission can be modeled as an absorbed, Comptonized blackbody in addition to Fe K$\alpha$ emission which can be attributed to relativistic disk reflection. We measure a persistent bolometric, unabsorbed luminosity of $L_{\mathrm{bol}}=7.03^{+0.04}_{-0.05}\times10^{36}\,\mathrm{erg\,s^{-1}}$, assuming a distance of 7\,kpc, corresponding to an Eddington ratio of $4.5\%$. This persistent luminosity combined with light curve analysis leads us to infer that the bursts were the result of pure He burning rather than mixed H/He burning. Time-resolved spectroscopy reveals that the \added{bolometric flux of the} first burst exhibits a double-peaked structure\replaced{which may be attributed to two ``sub-bursts" in quick succession, and}{, placing the source within a small population of accreting neutron stars which exhibit multiple-peaked type-I X-ray bursts. We find} that the second, brighter burst shows evidence for photospheric radius expansion (PRE)\replaced{. We show}{ and} that at its peak, this PRE event had an unabsorbed \added{bolometric} flux of $F_{\mathrm{peak}}=2.94^{+0.28}_{-0.26}\times10^{-8}\,\mathrm{erg\,cm^{-2}\,s^{-1}}$. This yields a new distance estimate of $d=9.0\pm0.5$\,kpc, assuming that this corresponds to the Eddington limit for pure He burning on the surface of a canonical neutron star. Additionally, we performed a detailed timing analysis which failed to find evidence for quasiperiodic oscillations or burst oscillations, and we place an upper limit of 16\% on the rms variability around 589\,Hz, the frequency at which oscillations have previously been reported.

\end{abstract}



\section{Introduction} \label{sec:intro}

Accretion onto compact objects produces an array of phenomena, the study of which can illuminate the nature of these objects and the process of accretion itself. Type-I X-ray bursts represent a particularly dramatic consequence of accretion, wherein the flux of an accreting neutron star (NS) is observed to increase by an order of magnitude or more in the span of only a few seconds or less. In a typical type-I burst, this fast rise in flux is followed by an exponential decay which can take anywhere from tens to thousands of seconds to return to a persistent flux level. Type-I X-ray bursts can provide a wealth of information about the compact stellar remnants from which they originate. The observed fast-rise exponential-decay behavior is thought to be due to runaway thermonuclear burning of accreted hydrogen and helium. This is possible only if the accretor has a surface onto which this material can accumulate in a thin shell which is unable to effectively cool during burning. Therefore, the detection of a type-I X-ray burst confirms the nature of an accreting source as a NS rather than a black hole. Additionally, in some cases type-I bursts allow for the measurement of other important parameters such as the distance to the source or the rotation period of the NS. For a recent review of type-I X-ray bursts, refer to \citet{Galloway2017}.

During a type-I X-ray burst, the source spectrum can be described by the addition of a blackbody component on top of whichever model best describes the persistent emission. Analyzing how the spectrum changes throughout a type-I burst is necessary in some cases to distinguish between changes in flux which are energy-independent and those which vary across different energy ranges, reflecting changes in the blackbody radius and temperature \citep{Tawara1984}. The latter case is often evidence for photospheric radius expansion (PRE). During PRE the blackbody which models the burst emission is shown to rapidly increase in radius while decreasing in temperature as the luminosity remains constant at the peak of the burst. This behavior has been attributed to radiation pressure lifting material from the NS surface upon reaching the Eddington limit. Due to the relatively narrow range of acceptable values for NS mass and radius, bursts which demonstrate PRE are often considered ``standard candles"  which can be used to estimate the distance to a bursting source \citep{vanParadijs1978,Lewin1984}.\added{ Some sources show significant scatter in peak luminosity across many PRE events however \citep{Kuulkers2003}. As such, these events may serve as approximate standard candles, and source properties such as disk inclination and the composition of burning material must be taken into account when calculating the Eddington luminosity.}

Additionally, timing analysis before, during, and after type-I bursts can reveal oscillatory behavior related to burning on the NS surface. Coherent high-frequency oscillations have been observed during some type-I bursts, most often during the decay in brightness \citep{Watts2012}. These oscillations have been attributed to brightness asymmetries induced by burning on the NS surface, implying that the frequency of burst oscillations corresponds to the rotational frequency of a NS. In some systems, quasi-periodic oscillations (QPOs) at mHz frequencies have also been observed in the persistent emission leading up to type-I bursts. This may indicate that as material accumulates, oscillatory nuclear burning occurs on the surface prior to the onset of the runaway burning which causes type-I bursts.

\grs\ is a low-mass X-ray binary (LMXB) which was first reported to exhibit type-I X-ray bursts in 1999 \citep{Cocchi1999}. The source resides near the Galactic Center, about $10^{\prime}$ from Sgr A* at $\alpha=17^{h}45^{m}02^{s}$, $\delta=-28^{\circ}54^{\prime}50^{\prime\prime}$ \citep{Muno2003}. \replaced{As an Atoll source, it accretes via Roche-lobe overflow}{As it accretes matter} from its main sequence companion, \replaced{transitioning}{it transitions} between a low-flux hard spectral state during which angular momentum transfer in the cold disk is inefficient, leading to low mass accretion rates, and a high-flux soft spectral state when the accretion disk \replaced{transitions to}{enters} a hot, efficiently accreting state \citep{Lasota2001}. \grs\ is a member of the very faint class of transients, meaning that it reaches a peak 2-10\,keV luminosity in the range of $\mathrm{10^{34-36}\,erg\,s^{-1}}$ \citep{Wijnands2005}. The first detection of \grs\ was made by the {\em GRANAT} satellite and reported in 1990, but was attributed to the nearby source 1E~1741.7-2850 \citep{Mandrou1990}. Later analysis resolved the source \citep{Sunyaev1990,Syunyaev1991}. \grs\ undergoes periods of outburst every $\sim2$\,years, with typical outbursts ranging in duration from 5 to 15 weeks, resulting in a duty cycle of $\sim10\%$. 

An investigation of the bursting behavior of \grs\ was presented by \citet{Trap2009}. The authors analyzed 15 type-I bursts observed during two periods of outburst in 2005 and 2007 by {\em INTEGRAL} JEM-X, {\em Swift} BAT and XRT, and {\em XMM-Newton}. The bursts had a typical recurrence time between 79\,ks and 2\,Ms. From this analysis they were able to infer a distance to the source of 7\,kpc, and they determined that the bursting behavior was consistent with pure He burning. During another period of outburst in 2013, the source was observed serendipitously during two observations by the Nuclear Spectroscopic Telescope Array (\nustar) \citep{Barriere2015}. One type-I X-ray burst lasting 800\,s was observed which was consistent with mixed H/He burning. It exhibited mild PRE, allowing the authors to infer a distance to the source which was consistent with the previous estimate of 7\,kpc. 

% Additionally the authors reported the possible detection of an absorption line in the burst spectrum at 5.5\,keV, which they tentatively attributed to gravitationally redshifted Chromium.

Again in April 2020, the source was observed by the XRT instrument aboard the Neil Gehrels {\em Swift} Observatory to be increasing in brightness, rising from $\mathrm{\sim10^{-2}\,count/s}$ to $\mathrm{\sim9\times10^{-2}\,count/s}$ over the course of two days, indicating that the source was entering a period of outburst \citep{Degenaar2020}. This detection was followed up by \nustar\ on May 7, 2020.

We report on the detection of two type-I X-ray bursts during the May 2020 \nustar\ observation of \grs\ in outburst, as well as spectral and timing analyses of the persistent emission. We begin by describing the observation and the methods of data reduction and analysis in Section \ref{sec:data}. Next, we present an analysis of the \nustar\ light curve, including modeling of each of the two type-I bursts in Section \ref{sec:lc}, the first of which is shown to peak twice. In Section \ref{sec:spectra} we present the persistent and burst spectra and demonstrate evidence for PRE during the second burst. For completeness, in Section \ref{sec:timing} we present an analysis of the timing features of both the persistent and burst emission, including a search for quasi-periodic oscillations and burst oscillations. We end with a discussion of our results in Section \ref{sec:discussion}.

\begin{figure*}
\begin{center}
\includegraphics[width=0.8\textwidth]{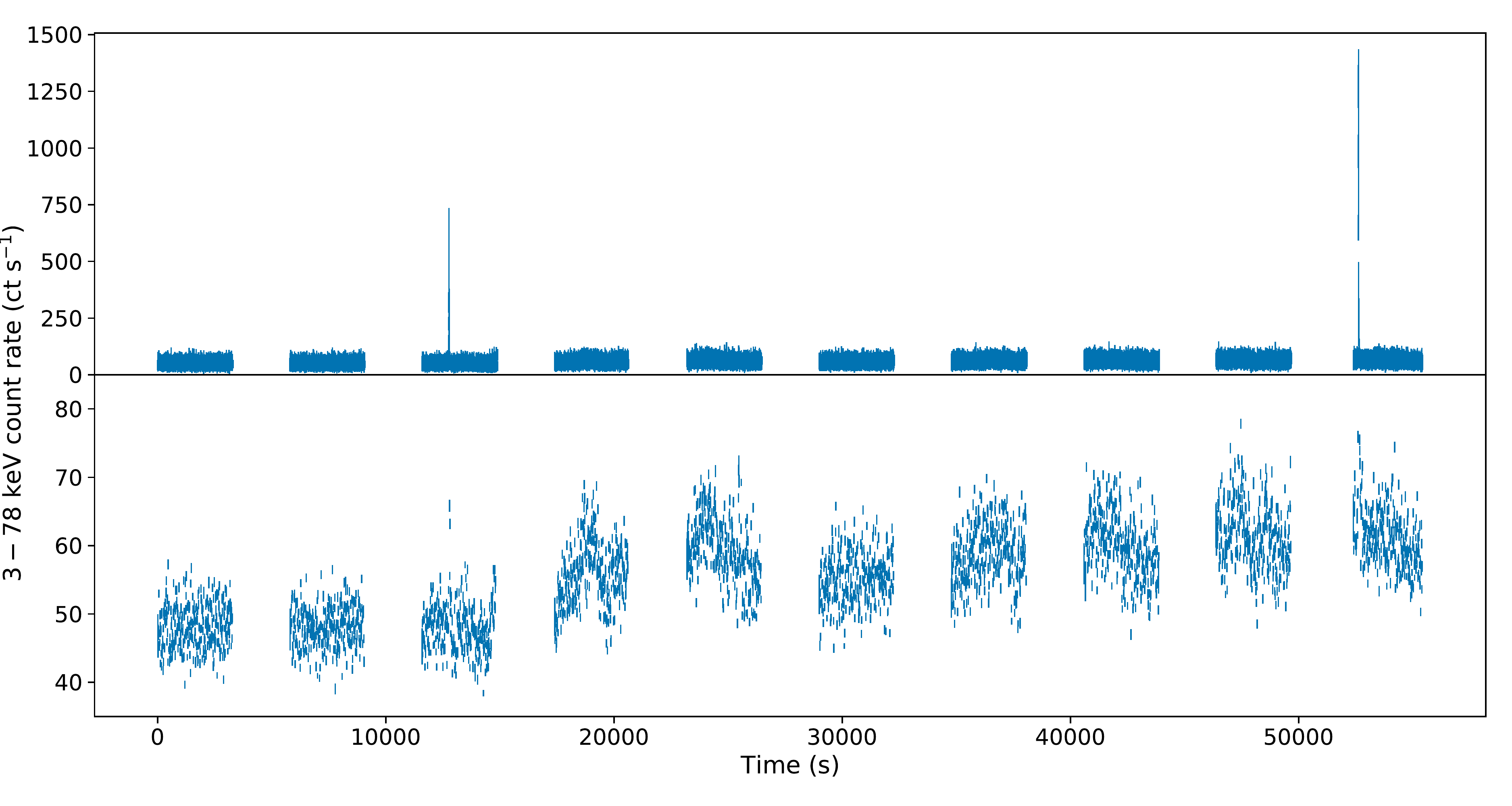}
\caption{The total, background subtracted, livetime corrected \nustar\ light curve is shown twice. The light curve in the top panel shows the entire dynamic range of the light curve and is binned with a resolution of 0.5\,s in order to illustrate the bursting behavior. The light curve in the bottom panel is the same as the top, but has been rebinned to a resolution of 10\,s and has been restricted along the y-axis in order to illustrate the behavior of the persistent emission.
\label{fig:lc}}
\end{center}
\end{figure*}

\section{Observation and Data Reduction} \label{sec:data}

\nustar, launched in June 2012, is the first high-energy focusing X-ray telescope \citep{Harrison2013}. It is composed of two focal planes, each paired with a set of focusing optics with a focal length of 10\,m. The focal planes, FPMA and FPMB, are each composed of 4 Cadmium Zinc Telluride (CZT) detectors attached to custom readout electronics.

\grs\ was observed by \nustar\ on May 7, 2020, for a total exposure time of ~28\,ks (OBSID 90601317002). We extracted the data products using NuSTARDAS version 2.0.0 and CALDB version 20200826. Photon arrival times were shifted to the barycenter of the solar system to eliminate residuals due to the orbital motion of \nustar. Instrumental effects on the photon arrival times due to changes in the temperature of the on-board clock were corrected using clockfile v110 generated on September 12, 2020. For the production of source spectra as well as cospectra, we used circular extraction regions with radius 100\,arcseconds, centered on the source using automatic centroid detection in DS9. To produce background spectra, we used extraction regions of radius 60\,arseconds, residing on the same detector as the source region but sufficiently removed from it so as to exclude source counts. Additionally, we found that in this case, the automated \replaced{correction}{pipeline overcorrected} for changes in the Multi Layer Insulation (MLI)\replaced{, described by \citet{Madsen2020}, was too aggressive and resulted}{ \citep{Madsen2020} resulting} in inconsistencies between FPMA and FPMB \replaced{at low energy ($<5$\,keV)}{below 5\,keV}. Therefore, we chose to revert to the previous FPMA ancillary response file.\footnote{For more information on how to determine when this is appropriate, refer to the  \href{http://nustarsoc.caltech.edu/NuSTAR_Public/NuSTAROperationSite/mli.php}{\nustar\ Science Operations Center homepage}.}

Timing analyses, such as the production of cospectra and calculation of epoch folding statistics, were performed using the Python package Stingray \citep{Huppenkothen2016}. Spectral modeling was performed using the X-ray spectral analysis package Xspec \citep[v12.11.1][]{Arnaud1996}. Spectra were binned using the variable binning algorithm described by \citet{Kaastra2016} which takes into account both the number of photons in a given energy bin as well as the average energy of the photons in that bin. We note that this binning procedure does not allow one to specify an exact energy range, meaning that the binned spectra may not extend all the way down to 3\,keV. We restricted our spectral analysis to an energy range of 3-40\,keV for persistent emission, and a range of 3-20\,keV for burst emission. At higher photon energies, background counts begin to contribute significantly to the overall spectra. All spectral fitting was performed using the Cash statistic \citep{Cash1979}, but throughout the paper we present chi-squared fit statistics in order to provide an idea of the quality of various fits using a formalism which is easy to interpret and which gives a clear comparison between models. For the purpose of readability, spectra shown in figures have been further rebinned such that each bin has at least 5-sigma significance. All light curves shown in this paper are the sum of the simultaneous light curves observed by FPMA and FPMB, and have been background subtracted and corrected for variations in the livetimes of the detectors. Errors quoted throughout this paper represent \replaced{1-sigma, or $68\%$,}{$90\%$} confidence regions unless otherwise stated.

\section{Light Curve Analysis} \label{sec:lc}

The total (3-78\,keV) \nustar\ light curve is shown in Figure \ref{fig:lc}, where the top panel shows the light curve binned in 0.5\,s increments in order to demonstrate the bursting behavior, and the bottom panel shows the light curve binned in 10\,s increments and restricted to count rates below $\mathrm{85\,count/s}$ in order to show the persistent flux level. Throughout the observation, \grs\ showed little variability aside from two dramatic increases in count rate occurring around 12700\,s and 52600\,s. Additionally, the persistent count rate appeared to increase gradually by about 30\% from about $\mathrm{50\,count/s}$ to about $\mathrm{65\,count/s}$. 

\begin{figure}
\begin{center}
\includegraphics[width=0.235\textwidth]{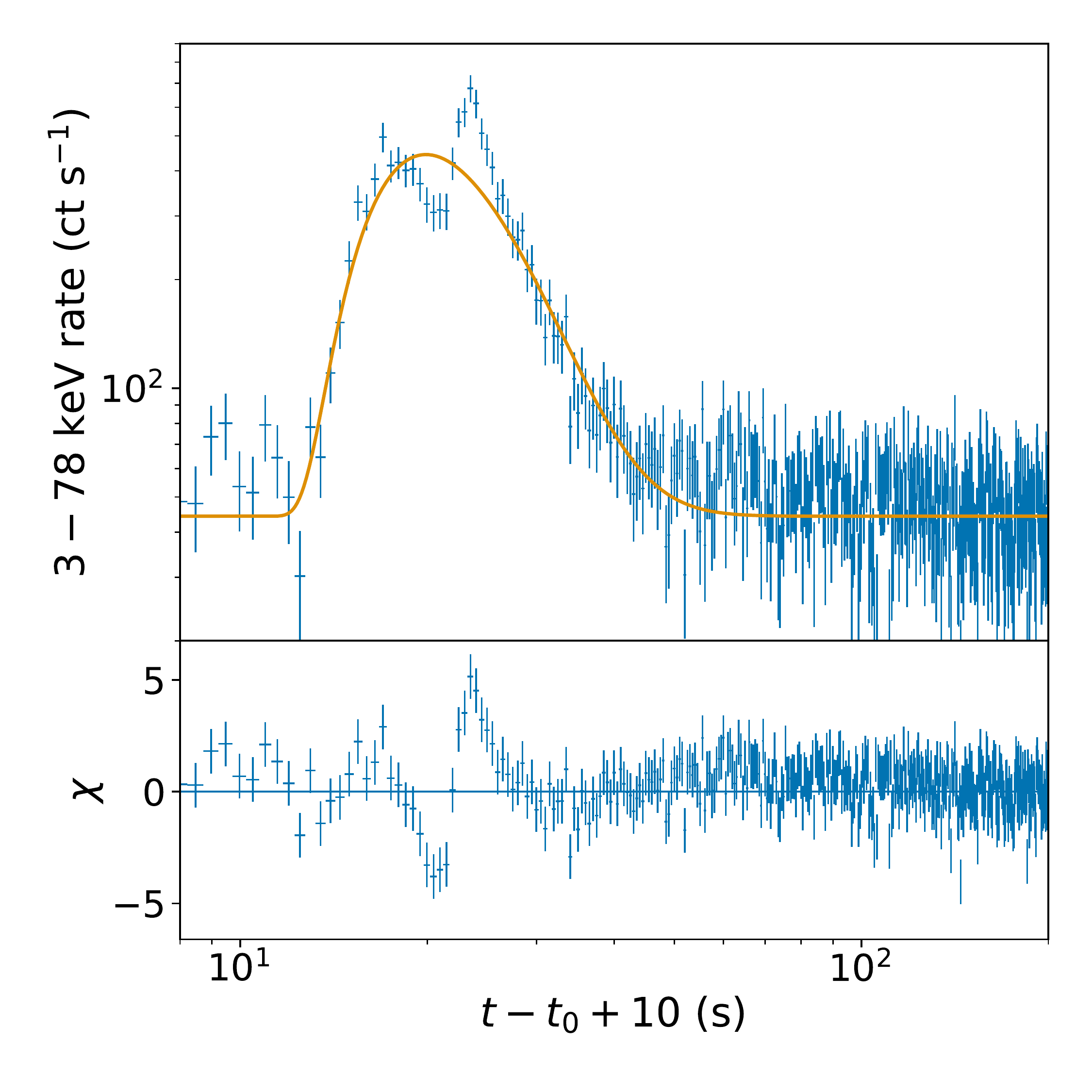}\includegraphics[width=0.235\textwidth]{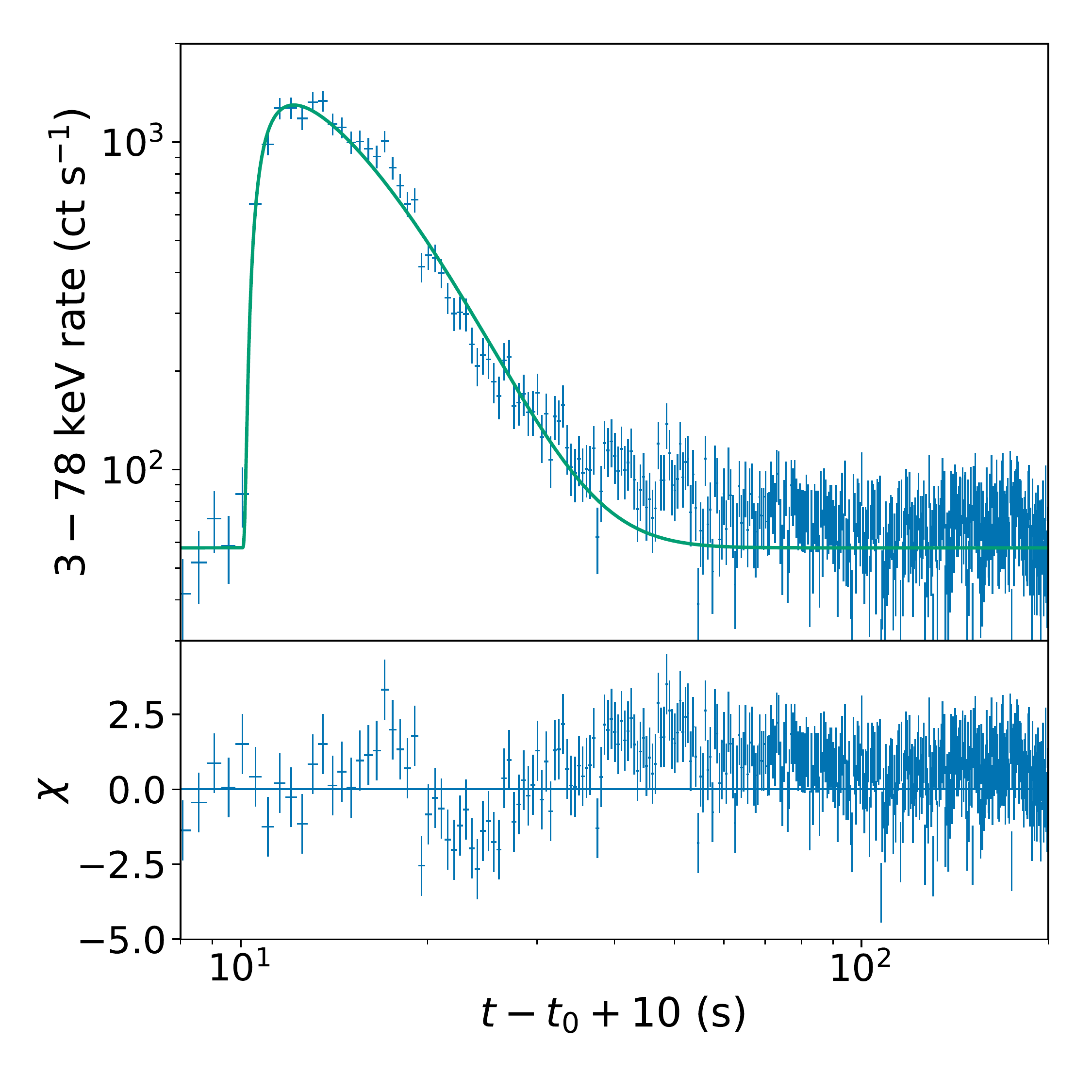}
\includegraphics[width=0.235\textwidth]{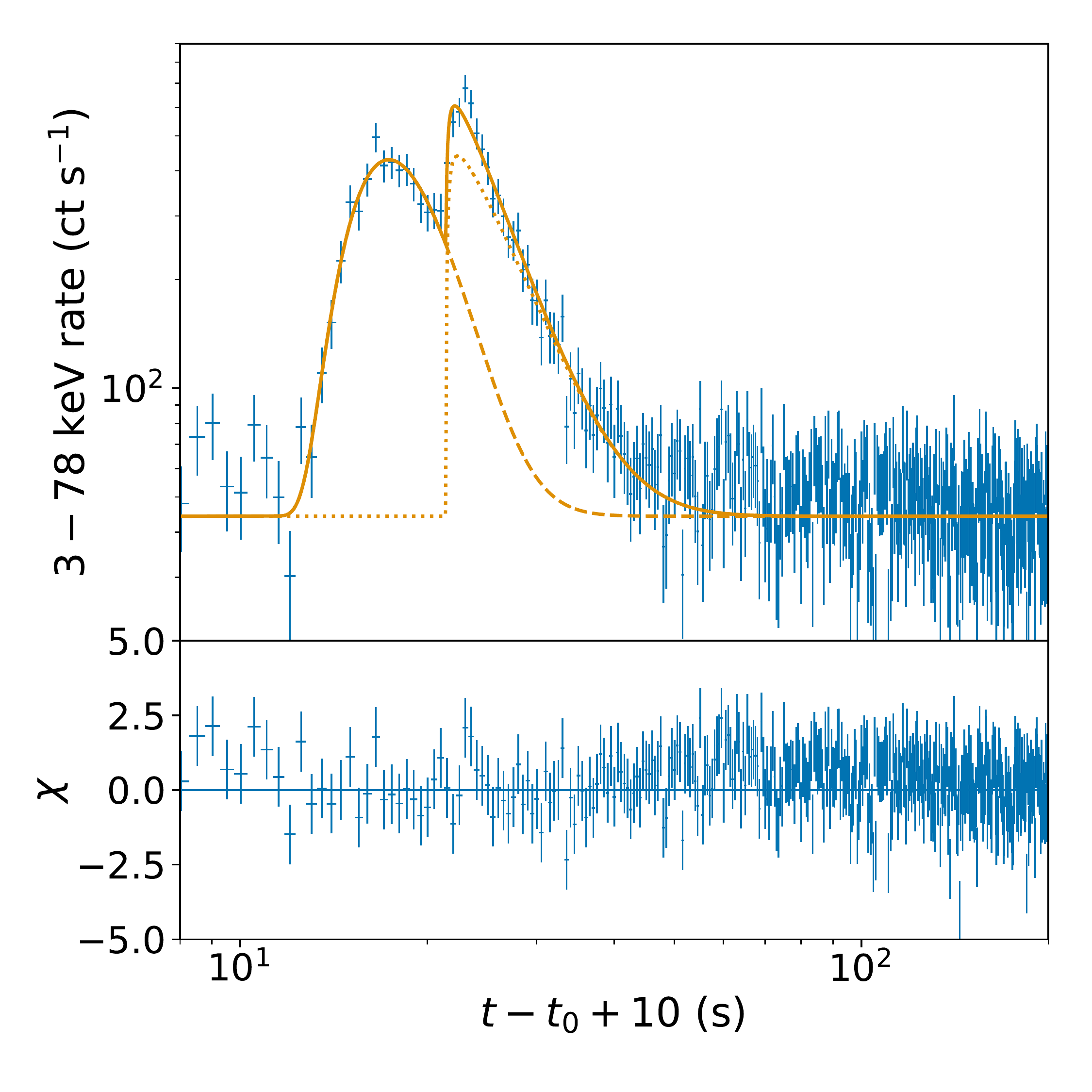}\includegraphics[width=0.235\textwidth]{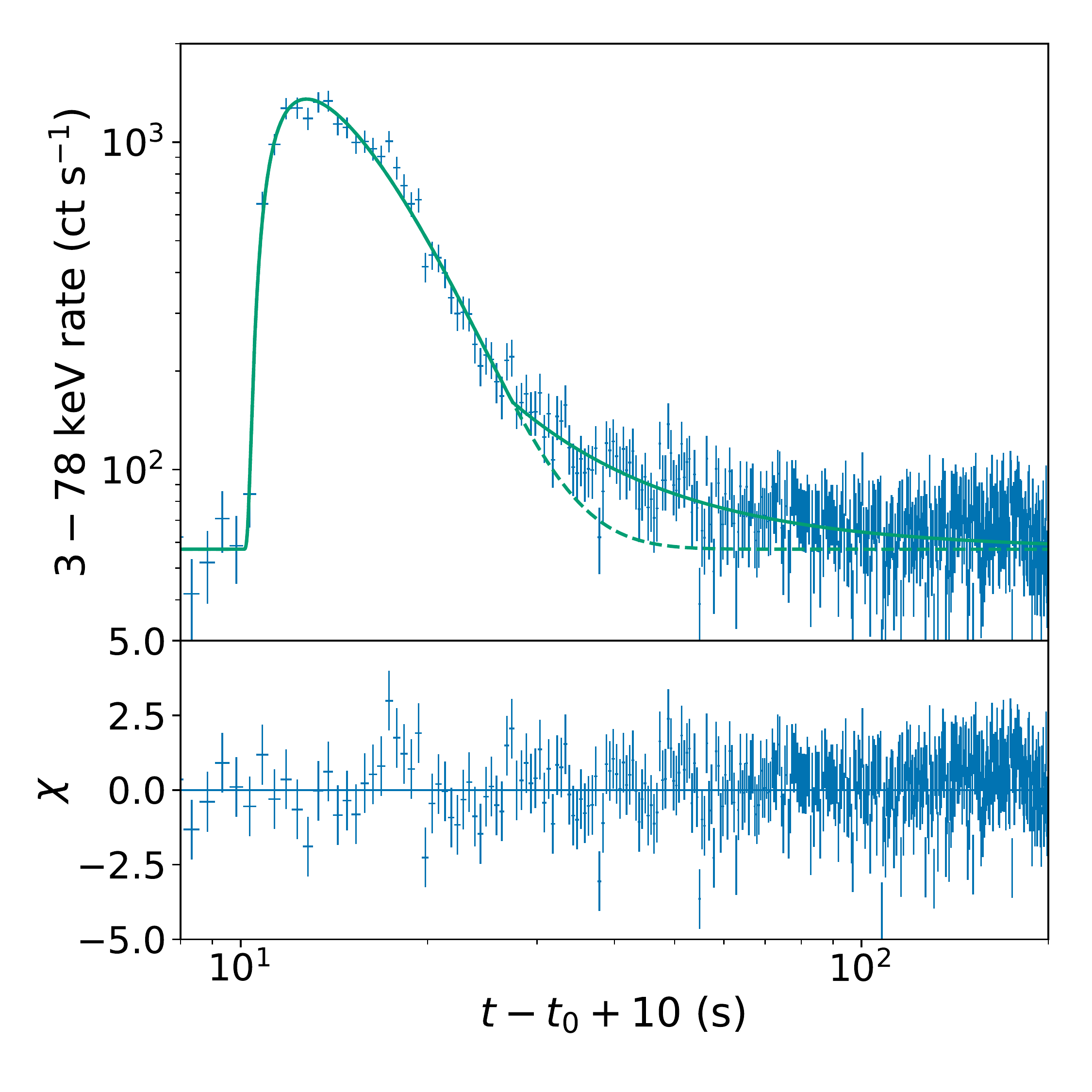}
\caption{Burst light curves along with their best-fit models and residuals. The observed light curves are shown in blue and have been plotted in logarithmic space, where the origin of the x-axis lies 10\,s prior to the onset of the corresponding burst, in order to more clearly illustrate the burst structure. The left two panels correspond to Burst 1 and the right two panels to Burst 2. The top panels show the bursts fit to a simple FRED model described by Equation \ref{eq:FRED} and shown by the solid lines. The bottom panels show the bursts fit to more complex models. Burst 1 is fit to a sum of two FRED sub-bursts, the first of which is plotted as a dashed line and second of which is dotted. Burst 2 is modeled by adding a power law component at late times and is described by Equation \ref{eq:FRED_pl}. The dashed line shows the FRED component and the solid line shows the total model. Both bursts are described more accurately by their respective augmented models rather than by a simple FRED model.
\label{fig:burst_fitting}}
\end{center}
\end{figure}

Due to their short duration and dramatic increase in count rate, the features at roughly 12700\,s and 52600\,s resemble type-I bursts. Indeed upon closer inspection, they exhibit the fast rise followed by an exponential decay which are characteristic of type-I bursts, confirming this classification. Figure \ref{fig:burst_fitting} shows the burst light curves for the entire 3-78\,keV \nustar\ band \deleted{as well as for several narrower energy bands, each} binned into 0.5\,s intervals. The bursts not only differ in maximum count rate, but in their structure as well. The first burst, hereafter Burst 1, appears to have a longer rise time and exhibits two clear peaks in count rate, while the second burst, hereafter Burst 2, exhibits a much faster rise and a single peak. 

\begin{deluxetable*}{ccccl}
\tablenum{1}
\tablecaption{light curve parameters for Bursts 1 and 2. Burst 1 is further broken down into sub-burst 1 and sub-burst 2. The top seven rows were determined via model fitting, while the bottom two rows were calculated based on the fitted parameters. The burst onset times, $t_0$, are given relative to the start of the observation, $t_\mathrm{ref}=\mathrm{MJD}\,58976.305553$.\label{tab:burst_param}}
\tablewidth{0pt}
\tablehead{
\colhead{Model Parameter}  & \colhead{sub-burst 1 Value} & \colhead{sub-burst 2 Value} & \colhead{Burst 2 Value} & \colhead{Units}}
\startdata
$t_0$       & $12752.5 \pm 2.3$     & $12763.9 \pm 0.2$ & $52614.7 \pm 0.1$   & s  \\
$\tau_{R}$  & $<52$\tablenotemark{$\dagger$}           & $<0.45$\tablenotemark{$\dagger$}   & $1.6 \pm 0.3$         & s  \\
$\tau_{D}$  & $2.3 \pm 1.3$         & $5.9 \pm 0.7$     & $4.9 \pm 0.3$         & s  \\
$A$         & $<1.4\times10^{6}$\tablenotemark{$\dagger$}  & $556 \pm 174$     & $4032 \pm 653$        & $\mathrm{count\,s^{-1}}$  \\
$t_{\mathrm{tail}}-t_0$ & \nodata   & \nodata           & $17.4 \pm 1.4$        & s  \\
$\gamma$    & \nodata               & \nodata           & $1.6 \pm 0.2$         & \nodata  \\
$C$         & \multicolumn{2}{c}{$44.3 \pm 0.3$}        & $57.1 \pm 0.3$        & $\mathrm{count\,s^{-1}}$ \\
\hline
$t_{\mathrm{peak}}-t_0$ & $7.3 \pm 2.4$     & $1.0 \pm 0.6$ & $2.8 \pm 0.3$     & s \\
$H$                     & $<3100$\tablenotemark{$\dagger$}    & $396 \pm 69$ & $1298 \pm 55$    & $\mathrm{count\,s^{-1}}$ \\
\enddata
\tablenotetext{\dagger}{$90\%$ upper confidence limit.}
\end{deluxetable*}

In order to better understand the structure of the two type-I bursts, we began by fitting the light curves to a simple Fast Rise Exponential Decay (FRED) model, given by

\begin{equation}
    f(t) = A\ exp \left[ -\frac{\tau_{R}}{t-t_{0}} - \frac{t-t_{0}}{\tau_{D}} \right] + C
    \label{eq:FRED}
\end{equation}

\noindent for $t>t_{0}$, where $t_{0}$ is the time at burst onset, $\tau_{R}$ and $\tau_{D}$ are the rise and decay times, respectively, $A$ is a factor which determines the height of the burst, and $C$ is the persistent count rate. From Equation \ref{eq:FRED} it can be seen that the burst peak occurs at
$t_{\mathrm{peak}}=\sqrt{\tau_{R}\tau_{D}}+t_{0}$. The height of the burst above the persistent contribution is therefore given by $H \equiv f(t_{\mathrm{peak}})-C=A\ exp[-2\sqrt{\frac{\tau_{R}}{\tau_{D}}}]$. The fits to the simple FRED model, including residuals, are shown in the top panels of Figure \ref{fig:burst_fitting}, in which the light curves are plotted in logarithmic space in order to better illustrate the structure of each burst. 

As expected, Burst 1 shows clear residuals around the peak of the FRED model indicating a double-peaked structure. We therefore proceeded to model Burst 1 as the sum of two ``sub-bursts" in quick succession, each modeled as a FRED burst. The fit to this model is shown in the bottom left panel of Figure \ref{fig:burst_fitting}. This results in a significantly better fit, eliminating the residuals around the two peaks. Fitting reveals that the first sub-burst has a significantly slower rise time than the second sub-burst but a somewhat faster decay time. It can also be seen in Figure \ref{fig:burst_fitting} that the two sub-bursts have comparable peak count rates. \deleted{From these characteristics we infer that the second sub-burst was not simply a small increase in flux during the decay of the first sub-burst but was in fact a separate sub-burst.} In Section \ref{sec:discussion} we discuss the \replaced{possible}{physical} mechanisms by which this double-peaked structure could be realized.

The second burst bears more resemblance to a typical type-I burst. The FRED model describes the shape of the burst well for early times, but excess emission can be seen in the tail of the burst. We therefore adopt a phenomenological model similar to the one used by \citet{Barriere2015}, in which the rise, peak, and beginning of the decay are modeled by a FRED curve, and the late-time emission is modeled as a power law decay rather than an exponential decay. This model can be written as

\begin{equation}
    f(t) = \begin{cases}
    A\ exp[-\frac{\tau_{R}}{t-t_{0}} - \frac{t-t_{0}}{\tau_{D}}] + C & \text{$t \leq t_{\mathrm{tail}}$} \\
    B\ (\frac{t-t_{0}}{t_{\mathrm{tail}}-t_{0}})^{-\gamma} + C & \text{$t>t_{\mathrm{tail}}$}
\end{cases}
\label{eq:FRED_pl}
\end{equation}

\noindent where $t_{\mathrm{tail}}$ is the time at which the emission transitions from an exponential tail to a power law tail, $\gamma$ is the power law index, and $B \equiv A\ exp[-\frac{\tau_{R}}{t_{\mathrm{tail}}-t_{0}} - \frac{{t_{\mathrm{tail}}-t_{0}}}{\tau_{D}}]$ such that the exponential tail and the power law tail have the same count rate at $t=t_{\mathrm{tail}}$. This model successfully describes the burst emission both at early and late times, as demonstrated by the bottom right panel of Figure \ref{fig:burst_fitting}. 

Table \ref{tab:burst_param} lists the values of the fitted and calculated parameters for Burst 1 and Burst 2. Both sub-burst 2 and Burst 2 show short rise times and similar decay times. The increase in count rate which is apparent from the the persistent light curve is reflected by the difference in the fitted values of $C$ between Burst 1 and Burst 2. The magnitude of the first sub-burst of Burst 1 is not precisely constrained due to degeneracy with other values and due to the limited number of constraining data points as compared to sub-burst 2, which dominates the light curve beginning about 8\,s after the onset of Burst 1.  With the onsets of each burst well-constrained, we are able to calculate the recurrence time between Burst 1 and Burst 2, $\Delta t = 39862.2 \pm 2.3\,\mathrm{s}$. This recurrence time is less than half the shortest recurrence time previously reported for \grs\ \citep{Trap2009}. Due to the orbital gaps in the light curve, however, it is possible that the two bursts were not consecutive.

\section{Spectral Analysis} \label{sec:spectra}

\begin{figure*}
\begin{center}
\includegraphics[width=0.45\textwidth]{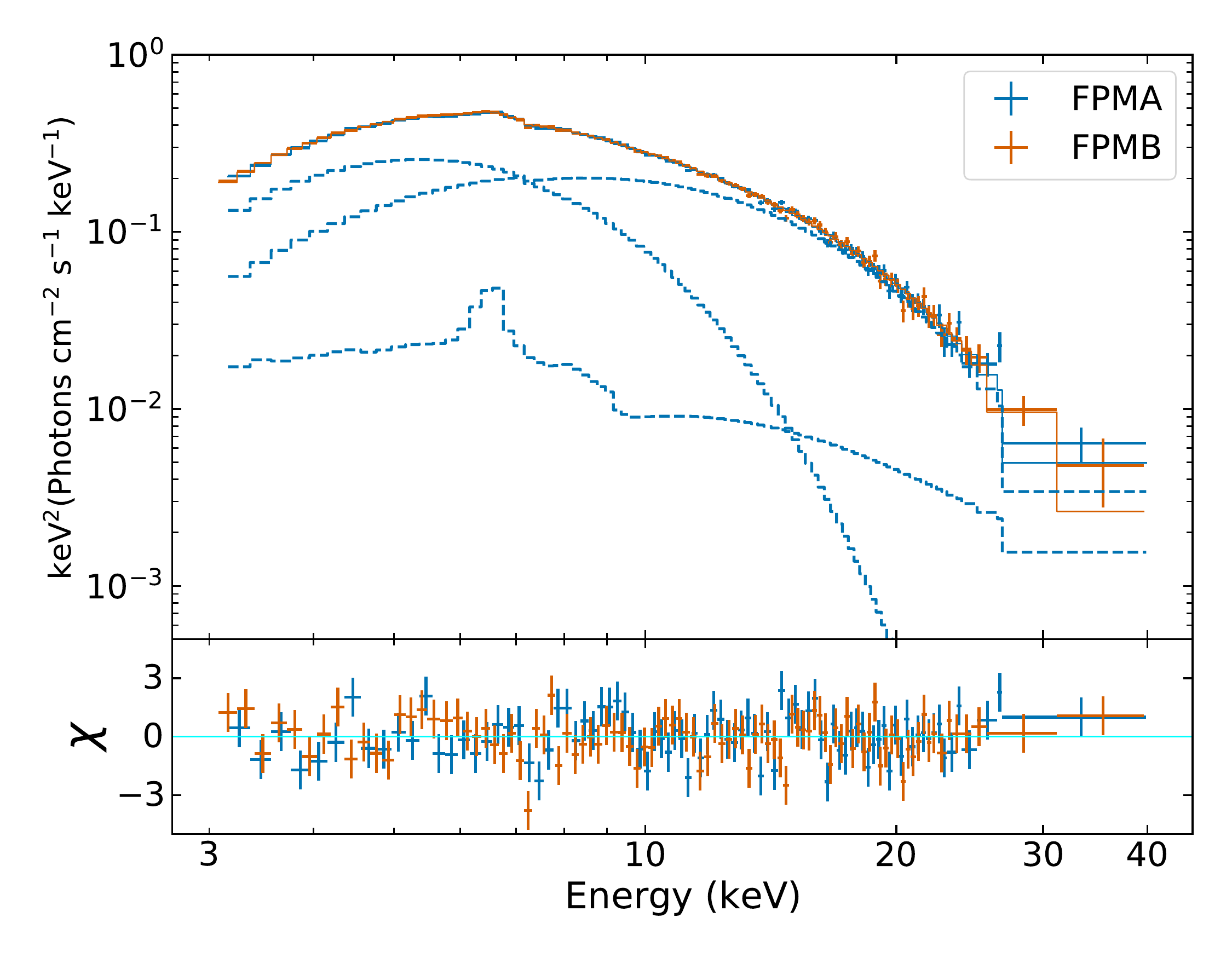}\includegraphics[width=0.45\textwidth]{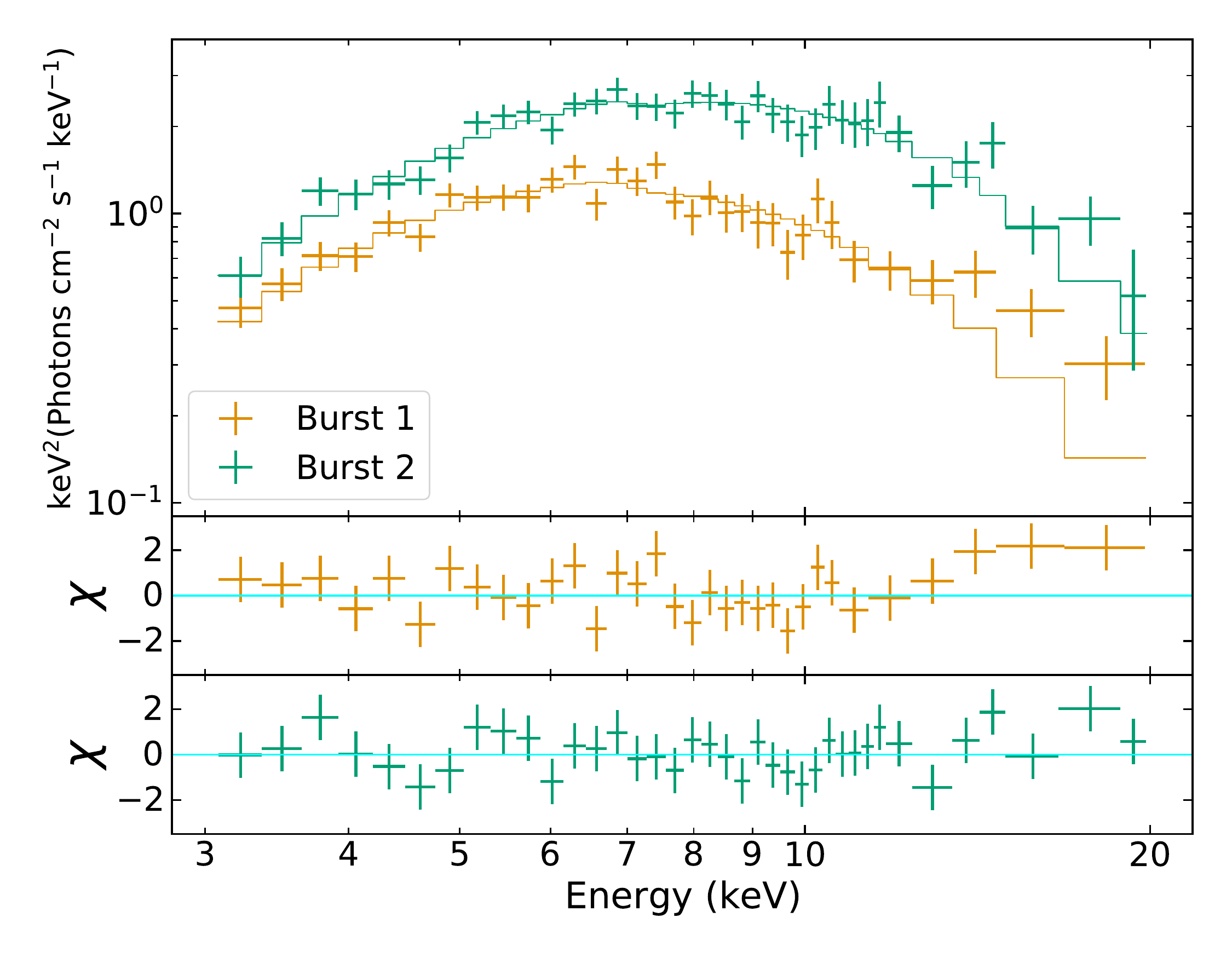}
\caption{Persistent and burst spectra and their best-fit models. The persistent spectra are shown on the left with FPMA in blue and FPMB in red. The solid lines show the best-fit reflection model, and the dashed lines shows the blackbody component, the Comptonized component, and the reflected cutoff power law component. The burst spectra are shown on the right with Burst 1 shown in orange and Burst 2 shown in green. For clarity, only the FPMA spectra are shown. The solid lines show the best-fit models which are dominated by the blackbody component which has been added to the persistent model.
\label{fig:spectra}}
\end{center}
\end{figure*}

\begin{deluxetable*}{ccrrr}
\tablenum{2}
\tablecaption{Spectral parameters determined by fitting the persistent spectra in Xspec. Model 1 corresponds to a Comptonized blackbody without disk reflection, Model 2 corresponds to relativistic disk reflection of an irradiating blackbody, and Model 3 corresponds to relativistic disk reflection of an irradiating cutoff power law. \label{tab:spectralPar}}
\tablewidth{0pt}
\tablehead{
\colhead{Component} & \colhead{Parameter} 							& \colhead{Model 1} & \colhead{Model 2} & \colhead{Model 3}}
\startdata
\noalign{\smallskip}
\code{tbabs}					& $N_{\rm H}$ ($10^{22}\,\mathrm{cm}^{-2}$) & $6.5^{+0.2}_{-0.3}$   &   $6.6^{+0.5}_{-0.2}$ &  $7.1^{+0.4}_{-0.3}$   \\
\noalign{\smallskip}
\hline
\noalign{\smallskip}
\multirow{2}{*}{\code{bbodyrad}}	& $kT$ (keV) 					    & $1.26 \pm 0.01$  & $1.23^{+0.01}_{-0.02}$	& $1.24 \pm 0.01$ \\
                                    & $(R_{\mathrm{km}}/d_{10})^{2}$	& $29 \pm 2$ 	   & $20 \pm 4$             & $26^{+1}_{-2}$      \\
\noalign{\smallskip}
\hline
\noalign{\smallskip}
\multirow{3}{*}{\code{nthcomp}}     & $\Gamma$ 			& $1.4^{+0.2}_{-0.1}$       & $2.0^{+0.2}_{-0.1}$       & $1.7^{+0.2}_{-0.1}$             \\
                                    & $kT_{e}$ (keV) 	& $2.65^{+0.09}_{-0.08}$   	& $2.87^{+0.20}_{-0.06}$    & $2.66^{+0.09}_{-0.06}$	\\
                                    & Norm ($10^{-3}$)  & $4.2^{+1.7}_{-1.2}$       & $10^{+3}_{-2}$            & $7.2^{+1.6}_{-0.9}$            \\
\noalign{\smallskip}
\hline
\noalign{\smallskip}
\multirow{3}{*}{\code{gaussian}}	& $E$ (keV) 							            & $6.48 \pm 0.07$           & \nodata   & \nodata   \\
                                    & $\sigma$ (keV) 						            & $0.24^{+0.10}_{-0.06}$    & \nodata	& \nodata   \\
  				                    & $K\ (10^{-4}\,\mathrm{photon\,cm^{-2}\,s^{-1})}$ 	& $4.5 \pm 1.0$ 	        & \nodata	& \nodata   \\
  				                    & $\mathrm{Equivalent\ Width\ (eV)}$ 	            & $39^{+7}_{-4}$ 	        & \nodata	& \nodata   \\
\noalign{\smallskip}
\hline
\noalign{\smallskip}
\multirow{6}{*}{\code{relxill}/\code{relxillNS}}    & $q$                                 & \nodata   & $3.5$\tablenotemark{$\dagger$} & $3.0$\tablenotemark{$\dagger$} \\
                                                    & Inclination ($ ^{\circ}$)         & \nodata   & $13^{+16}_{-7}$      &  $17^{+13}_{-11}$ \\
                                                    & $R_\mathrm{in}$ ($R_\mathrm{g}$)	& \nodata   & $61^{+74}_{-15}$          &  $44^{+78}_{-18}$ \\
  				                                    & $\log\xi\ (\mathrm{erg\,cm\,s^{-1}})$ 				        & \nodata   & $2.59^{+0.06}_{-0.23}$    &  $2.98^{+0.05}_{-0.40}$ \\
  				                                    & $A_{\mathrm{Fe}}$ (Solar)         & \nodata   & $3.0 \pm 1.9$ 		    &  $2.1^{+2.1}_{-1.0}$ \\
  				                                    & Norm ($10^{-4}$)		            & \nodata	& $1.1^{+1.6}_{-0.8}$ 	    &  $3.1^{+0.9}_{-0.8}$ \\
\noalign{\smallskip}
\hline
\hline
\noalign{\smallskip}
\multicolumn{2}{c}{$\chi^{2}/\mathrm{d.o.f.}$}  & 293.5/246    & 284.5/244    & 274.9/244 \\
\noalign{\smallskip}
\enddata
\tablenotetext{\dagger}{The values of the disk emissivity index were fixed during fitting.}
\end{deluxetable*}

\begin{deluxetable}{ccc}
\tablenum{3}
\tablecaption{Physical quantities calculated for the persistent emission, based on the non-reflection spectral model.\label{tab:flux}}
\tablewidth{0pt}
\tablehead{
\colhead{Quantity} & \colhead{Value} & \colhead{Units}}
\startdata
$F_{\mathrm{bol}}$\tablenotemark{a}     & $1.20 \pm 0.01$ & $\mathrm{10^{-9}\ erg\ cm^{-2}\ s^{-1}}$        \\
\noalign{\smallskip}
$L_{\mathrm{bol}}$\tablenotemark{b}     & $7.03^{+0.04}_{-0.05}$ & $\mathrm{10^{36}\ erg\ s^{-1}}$   \\
\noalign{\smallskip}
$\dot{m}$\tablenotemark{c}    & \comment{$4.03^{+0.04}_{-0.03}$} $3.9$ & $10^{3}\ \mathrm{g\ cm^{-2}\ s^{-1}}$  \\
\noalign{\smallskip}
$\dot{m}/\dot{m}_{\mathrm{Edd}}$\tablenotemark{d}   & $4.5$  & $\%$ \\
\noalign{\smallskip}
$\dot{M}$\tablenotemark{e}   & $7.8$ & $10^{-10}\ M_{\odot}\ \mathrm{yr^{-1}}$ \\
\noalign{\smallskip}
\enddata
\tablenotetext{a}{\footnotesize Bolometric (0.1-100\,keV) unabsorbed flux.}
\tablenotetext{b}{\footnotesize Bolometric unabsorbed luminosity assuming a distance of 7\,kpc.}
\tablenotetext{c}{\footnotesize Mass accretion rate per unit surface area as determined by Equation \ref{eq:m_dot}.}
\tablenotetext{d}{\footnotesize Ratio of the mass accretion rate to the Eddington limited rate, $\dot{m}_{\mathrm{Edd}}=8.8\times10^{4}\,\mathrm{g\,cm^{-2}\,s^{-1}}$, assuming $M=1.4\,M_{\odot}$ and $R_{\mathrm{NS}}=10$\,km.}
\tablenotetext{e}{\footnotesize Mass accretion rate assuming a NS radius of 10\,km.}
\end{deluxetable}

In order to better understand the bursting behavior of \grs, it is necessary to characterize both the persistent and burst spectra, as well as to understand how the spectrum changes during each burst. We began by producing good time intervals (GTI) for the persistent emission as well as for each of the two type-I bursts. For the first burst, we defined the GTI to begin at the onset of the first sub-burst and to end after 5 decay times ($\tau_D$) had elapsed following the onset of the second sub-burst. This results in an interval with length 41\,s. For the second burst, we similarly defined the GTI to begin at the onset of the burst and to end after 5 decay times ($\tau_D$) had elapsed, resulting in an interval with length 24\,s. The GTI for persistent emission was defined such that it excluded the bursts, with a 100\,s buffer prior to the onset of each burst, and a 500\,s buffer following the end of each burst's GTI, such that in total 1265\,s of the observation was excluded. Using these good time intervals, we were able to extract the persistent and burst spectra.\added{ When extracting burst spectra, we loosened the event filters by specifying the status expression ``STATUS=\code{b0000xx000xxxx000}" when running \code{nupipeline}. We therefore were able to avoid removing source photons which would otherwise be mistaken for spurious events at such high count rates.}

\subsection{Persistent Emission} \label{subsec:persistent_spectra}
The persistent spectrum is described well by the standard physical picture of a NS surface or boundary layer emitting blackbody radiation which is Compton upscattered into a powerlaw-like component by a hot corona. In Xspec, we represented the blackbody emission using \code{bbodyrad} and the upscattered emission using \code{nthcomp} \citep{Zdziarski1996,Zycki1999}. The seed photon temperature for \code{nthcomp} was tied to the temperature of the blackbody. An absorbing column, represented by \code{tbabs} was also applied to the sum of the model components, with molecular abundances described by \citet{Wilms2000} and cross-sections described by \citet{Verner1996}. Fitting with this model resulted in a reduced Chi-squared statistic of $\chi^{2}_{\nu} = 1.64$ ($\chi^{2}=408.63$; $\mathrm{d.o.f.}=249$). We observed a clear Fe K$\alpha$ emission feature in the resulting residuals, so we also included a Gaussian component at $\approx 6.5\,\mathrm{keV}$. The addition of this component improved the fit significantly, bringing the reduced Chi-squared down to $\chi^{2}_{\nu} = 1.19$ ($\chi^{2}=293.5$; $\mathrm{d.o.f.}=246$), confirming that the component is necessary to accurately describe the observed spectrum.

Additionally, we found that the Fe line emission could be accounted for by replacing the Gaussian component with a relativistic disk reflection model. We found that the spectrum could be described well by the reflection off an accretion disk of a cutoff power law approximating the \code{nthcomp} component using the \code{relxill}\footnote{We note that although the alternative reflection model \code{relxillCp} models the incident power law using \code{nthcomp}, that model assumes that the Comptonized blackbody has a disk geometry with fixed temperature of $kT_{e}=0.05\,\mathrm{keV}$. In order to maintain consistency between model components we therefore chose to use \code{relxill} rather than \code{relxillCp}} model \citep{Dauser2014,Garcia2014}, or by the reflection of the blackbody component using a modified version of \code{relxill}, \code{relxillNS} (Garc\'ia et al., subm.). Although reflection of the blackbody using \code{relxillNS} can account for the line emission, it cannot account for the hard emission on its own, and the \code{nthcomp} component is still necessary to accurately model the spectrum up to high photon energies.

For both of our reflection models, the disk emissivity index was assumed to be constant throughout the disk, the disk density was fixed at $n=10^{15}\,\mathrm{cm^{-3}}$, the spin parameter was fixed at $a=0$, and the reflection fraction was fixed at a value of $-1$, such that only the reflected component was modeled by \code{relxill} and \code{relxillNS} (in other words, we added the reflected component to the non-reflection model described above, rather than replacing the direct blackbody and Comptonized emission with the \code{relxill} models). In the case of blackbody reflection, the emissivity index of the disk was fixed to $q=3.5$, corresponding to irradiation of the disk by the NS surface or boundary layer \citep{Wilkins2018}, and the temperature of the incident blackbody was tied to that of the \code{bbodyrad} component. For the case of power law reflection, we fixed the emissivity index at a value of $q=3$ corresponding to a disk corona geometry. We tied the \code{relxill} power law index to that of \code{nthcomp}, and we related the \code{relxill} cutoff energy to the \code{nthcomp} electron temperature as $E_\mathrm{cut}=3kT_{e} - 2\,\mathrm{keV}$ such that the shape of the incident power law roughly approximated the component described by \code{nthcomp}.

The best-fit parameters for all three models are listed in Table \ref{tab:spectralPar}, where the non-reflection model is referred to as Model 1, the reflected blackbody (\code{relxillNS}) is referred to as Model 2, and the reflected cutoff power law (\code{relxill}) is referred to as Model 3. We find that the three models provide a similarly good fit: the reduced chi-squared statistics for Models 1, 2, and 3 are $\chi^{2}_{\nu}=1.19\ (\mathrm{d.o.f}=246)$, $1.17\ (\mathrm{d.o.f}=244)$, and $1.13\ (\mathrm{d.o.f}=244)$, respectively. The observed spectra as well as the residuals and components of the Model 3 are shown in the left panel of Figure \ref{fig:spectra}. The three models are largely consistent with one another, showing little change between their shared components. Additionally, the spectral parameters are consistent with previous measurements \citep{Barriere2015} with somewhat lower column density and blackbody temperature. 

From the non-reflection model, we find an apparent blackbody radius of $R=3.8 \pm 0.1\ \mathrm{km}$ assuming a distance to the source of 7\,kpc. Given that the gravitational redshift on the surface of the NS, given by $1+z=(1-2GM/(c^{2}R_{\mathrm{NS}}))^{-1/2}$, is significant, the radius measured by fitting to a blackbody is larger than the radius in the emitting frame by a factor of $(1+z)^{3/2}$. Assuming $M=1.4\,M_{\odot}$ and $R_{\mathrm{NS}}=10$\,km, we get $1+z=1.31$. This yields an actual blackbody radius of 2.5\,km, consistent with accretion onto a small band rather than spherical accretion onto the entire surface of the NS. The relativistic reflection models provide similar values for the ionization of the disk, $\xi\sim10^{3}\ \mathrm{erg\,cm\,s^{-1}}$, as well as the iron abundance, $A_\mathrm{Fe}$, which is loosely constrained at about 3 times the Solar value. The iron abundance, however, depends strongly on the disk density which cannot be varied using \code{relxill} and which we could not constrain with \code{relxillNS}. We therefore advise against interpreting the iron abundance as the true physical value. Both \code{relxill} and \code{relxillNS} yield a low inclination of $\sim15^{\circ}$, and an inner disk radius consistent with $\sim50\,R_{g}$, corresponding to about $100$\,km for a NS mass of $M=1.4\,M_{\odot}$.

Having modeled the persistent emission, we were able to constrain the persistent flux and corresponding mass accretion rate. From \citet{Galloway2008}, the bolometric luminosity, $L_{\mathrm{bol}}$, is related to the accretion rate per unit area, $\dot{m}$, by $L_{\mathrm{bol}}=4\pi R_{\mathrm{NS}}^{2}\dot{m}(GM/R_{\mathrm{NS}})(1+z)^{-1}$. Solving for $\dot{m}$, as in \citet{Barriere2015}, gives

\begin{equation}
\begin{aligned}
    \dot{m}= & 3280\times\left(\frac{F_{\mathrm{bol}}}{10^{-9}\,\mathrm{erg\,cm^{-2}\,s^{-1}}}\right)\left(\frac{M}{1.4\,M_{\odot}}\right)^{-1}\\
    & \times \left(\frac{R_{\mathrm{NS}}}{10\,\mathrm{km}}\right)^{-1}\left(\frac{d}{7\,\mathrm{km}}\right)^{2}\left(\frac{1+z}{1.31}\right)\,\mathrm{g\,cm^{-2}\,s^{-1}}
\end{aligned}
\label{eq:m_dot}
\end{equation}

\noindent where $F_{\mathrm{bol}}$ is the bolometric flux, $M$ and $R_{\mathrm{NS}}$ are the NS mass and radius, respectively, $d$ is the distance to the source, and $z$ is the gravitational redshift at the surface of the NS. The values which we measured for the bolometric flux and the resulting calculated values of bolometric luminosity and accretion rate \added{,assuming a distance of $d=7\,\mathrm{kpc}$,} are shown in Table \ref{tab:flux}.
 
\subsection{Burst Emission} \label{subsec:burst_spectra}

\begin{figure*}
\begin{center}
\includegraphics[width=0.45\textwidth]{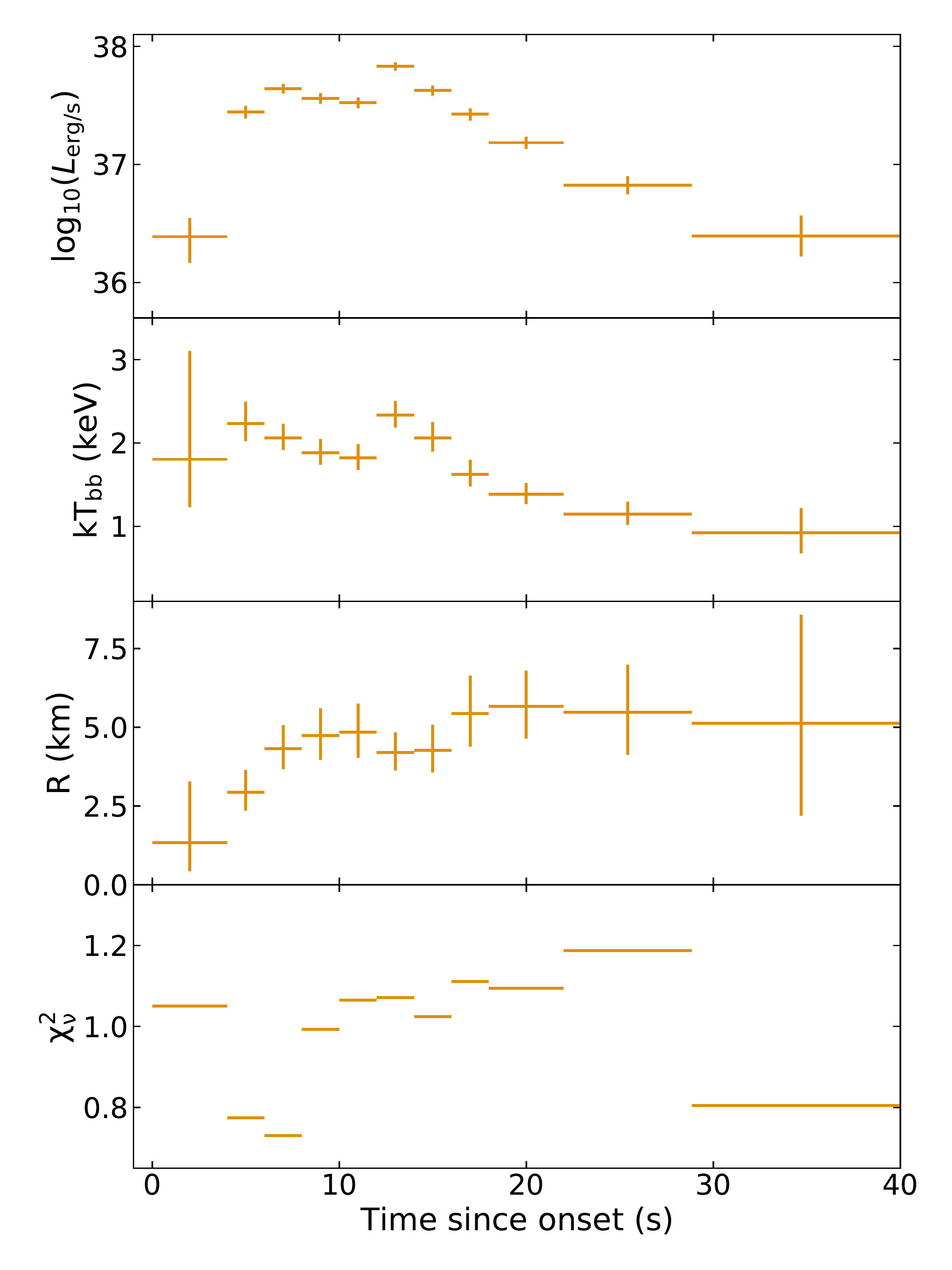}\includegraphics[width=0.45\textwidth]{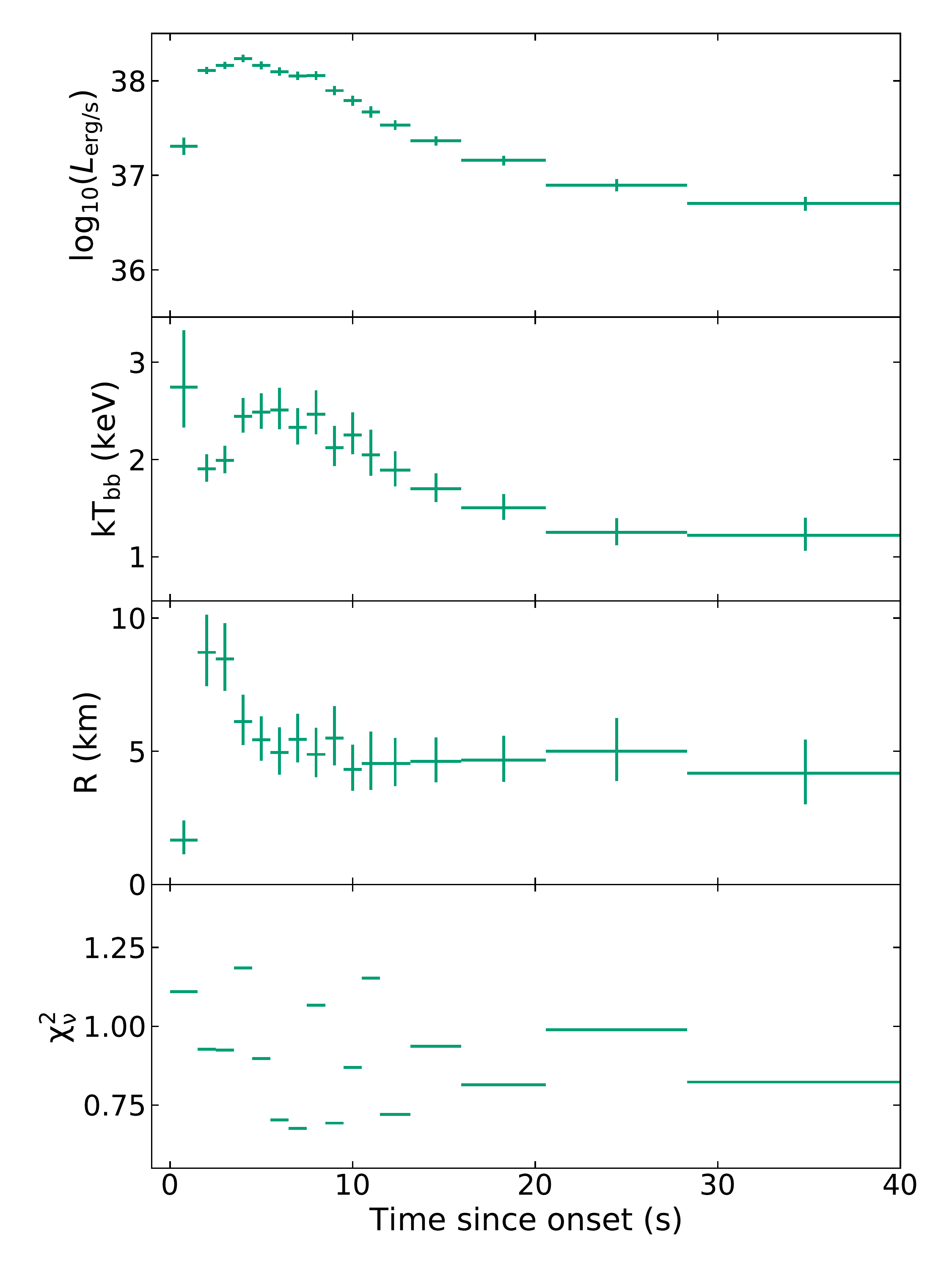}
\caption{Evolution of the blackbody component which describes the emission for each type-I X-ray burst. The apparent unabsorbed bolometric luminosity, temperature, and radius (assuming a distance of 7\,kpc) of the blackbody are shown for Burst 1 on the left in orange and for Burst 2 on the right in green. The double-peaked structure of Burst 1 is observed in the intrinsic luminosity, confirming that it is not the result of the spectrum dropping below the \nustar\ bandpass. Burst 2 exhibits a PRE ``touchdown" event a few seconds after burst onset in which the blackbody radius contracts rapidly while the temperature increases at constant luminosity.
\label{fig:PRE}}
\end{center}
\end{figure*}

The burst spectra were modeled by freezing the persistent model (for simplicity, we used the non-reflection model) and adding an additional blackbody component which was allowed to vary in both temperature and normalization.\footnote{We also tried allowing the total normalization of the persistent model to vary by introducing a multiplicative factor\added{, following the approach of \citet{Worpel2013,Worpel2015}}. For Burst 1 we found that this factor was difficult to constrain, and for Burst 2 the factor was consistent with unity and its inclusion resulted in a negligible improvement to the fit statistic. We similarly found that adding a second blackbody component did little to improve the fits and was difficult to constrain.} Rather than pinning the spectral parameters of the persistent model to the values determined for the full observation, we determined the persistent model parameters relevant to each burst using spectra extracted near the time of that burst, excluding the interval of variability between $\sim15$\,ks and $\sim35$\,ks. 

\deleted{The right panel of Figure \ref{fig:spectra} shows the spectra during both bursts as observed by FPMA. We note that FPMB has been omitted from the figure for clarity but shows no significant deviations from the spectrum observed by FPMA. As expected, the blackbody temperature inferred from the spectrum of Burst 2 is higher than that of Burst 1, and the blackbody temperatures during the bursts are significantly higher than that observed in the persistent state. During Burst 1, a temperature of $kT=1.88^{+0.06}_{-0.05}\,\mathrm{keV}$ is observed, and during Burst 2, a temperature of $kT=2.08 \pm 0.05\,\mathrm{keV}$ is observed. These fits represent the mean spectra during each of the bursts however, and the spectrum is expected to evolve significantly throughout the burst, even on very short timescales. Therefore, we broke the observation up further in order to perform a time-resolved spectroscopic analysis for each type-I X-ray burst.}

For each burst, we produced GTIs of varying length, beginning at the burst onset and ending after several decay times. The GTI lengths were set such that each bin contained a roughly equal number of counts, and so that we could achieve good temporal resolution during the first several seconds of the burst while also maintaining enough photons per time bin to enable robust spectral fitting. For each of these intervals, the spectrum was again extracted and fit to the persistent-plus-blackbody model described above, allowing the blackbody temperature and normalization to vary. The resulting parameters, including the temperature, the unabsorbed bolometric luminosity, and radius of the blackbody (assuming a distance of 7\,kpc) are shown in Figure \ref{fig:PRE}. Each data point corresponds to a GTI for which spectra were analyzed.

Time-resolved spectroscopy reveals that the double-peaked structure observed in the count rate during Burst 1 is also visible in the luminosity and temperature evolution of the burst. The first sub-burst reached a blackbody temperature of $kT_{\mathrm{bb}}=2.2\,\mathrm{keV}$ and a luminosity of $L_{\mathrm{bol}}=4.4\times 10^{37}\,\mathrm{erg\,s^{-1}}$, while the second sub-burst reached a blackbody temperature of $kT_{\mathrm{bb}}=2.3\,\mathrm{keV}$ and a luminosity of $L_{\mathrm{bol}}=6.8\times 10^{37}\,\mathrm{erg\,s^{-1}}$. This analysis confirms that the double-peaked structure is intrinsic to the source. That is, it is not the result of an increase in the fraction of incident photons energies falling below \nustar's lower energy limit of 3\,keV due to a softening of the spectrum at a constant luminosity. The available data do not suggest that the NS underwent PRE during either of the sub-bursts which make up Burst 1, but our analysis is limited by our inability to resolve these short sub-bursts into smaller time bins due to insufficient counts.

We showed in Section \ref{sec:lc} that Burst 2 exhibited a very different structure compared Burst 1. This is borne out in our time-resolved spectroscopic analysis as well. The blackbody luminosity shows a single peak, reaching a bolometric flux of $F_{\mathrm{peak}}=2.94^{+0.28}_{-0.26}\times 10^{-8}\,\mathrm{erg\,cm^{-2}\,s^{-1}}$ ($L_{\mathrm{peak}}=1.7\times 10^{38}\,\mathrm{erg\,s^{-1}}$, assuming a distance of 7\,kpc) within a few seconds of the onset of the burst, and a long decay time. On the other hand, the apparent blackbody temperature and radius show variability during the luminosity peak, with the temperature decreasing rapidly from $2.7$\,keV to $1.9$\,keV in the first second of the burst, followed by an increase up to $2.5$\,keV only a few seconds later. This temperature evolution is accompanied by a rapid expansion and contraction of the apparent blackbody radius, which reaches a maximum of $8.7$\,km before contracting to about $5$\,km. \added{We note that the observed values of the blackbody radius and temperature are under- and overestimates, respectively, of their actual physical values. Scattering of the blackbody emission in the NS atmosphere results in a harder emergent spectrum. The overall effect of this scattering is to introduce a color correction factor, $f_\mathrm{c}$, to the effective temperature so that $T_\mathrm{eff} \propto T_\mathrm{obs}f_\mathrm{c}^{-1}$, where $T_\mathrm{obs}$ is the temperature determined via modeling of the observed spectrum. Similarly, the effective radius is related to the observed radius by $R_\mathrm{eff} \propto R_\mathrm{obs}f_\mathrm{c}^{2}$. The color correction factor may vary between 1.4 and 1.7 for luminosities approaching the Eddington limit \citep{Suleimanov2011}. Therefore at the peak of Burst 2, the blackbody temperature may be a factor of $\sim0.6$ lower than that observed, and the blackbody radius a factor of $\sim3$ larger.}

\replaced{This}{The} behavior \added{described above} is the signature of a ``touchdown" event wherein the photosphere undergoes expansion due to radiation pressure, then rapidly falls back down to the surface such that luminosity remains constant while the contraction causes the temperature to increase. We therefore conclude that the NS underwent PRE during Burst 2. Because this phenomenon is generally attributed to the balance of radiation pressure and gravitational force at the Eddington limit ($L_{\mathrm{Edd}}\sim 2\times 10^{38}\,\mathrm{erg\,s^{-1}}$ for a NS with typical mass $M=1.4\ M_{\odot}$), it is often assumed that PRE allows for the measurement of the source distance. We defer estimation of the distance and further discussion until Section \ref{sec:discussion}.

\section{Timing Analysis} \label{sec:timing}

We searched both the persistent and burst emission for timing features including quasi-periodic and coherent oscillations. We began by producing cross-spectra between FPMA and FPMB for $113 \times 256$\,s intervals included in the persistent GTI described in Section \ref{sec:spectra}. We analyzed the cospectrum, the real part of the cross-spectrum, due to its advantages described by \citet{Bachetti2015}, namely the fact that it eliminates contributions to the power spectrum introduced by Poisson noise and dead time, the latter of which can be significant even for moderate \nustar\ count rates. In order to produce the cospectra, we used Stingray to produce light curves using only events within the $100^{\prime\prime}$ source extraction region and with energy between $3$ and $78$\,keV. We binned the light curves with a resolution of $2048^{-1}$\,s, and ensured that the light curves were simultaneous between FPMA and FPMB. 

We inspected the resulting dynamical cospectrum as well as the averaged cospectrum to look for coherent signals, particularly around the previously detected signal at $589$\,Hz \citep{Strohmayer1997}. There is no indication of such a feature, and the cospectrum of the persistent emission does not appear to deviate from a simple power law. We confirmed this visual inspection by performing a chi-squared fit to a power law model, which yielded a reduced chi-squared of $\chi^{2}/\mathrm{d.o.f.}=0.98$. Nonetheless, we performed an automated search for QPO features by fitting the average cospectrum to a power law with an added Lorentzian feature, such that the total model is given by

\begin{equation}\label{eq:Lorentz_pl}
    f(\nu) = A \nu^{\alpha} + B \left( \frac{1}{\pi} \right) \left[ \frac{\gamma}{(\nu-\nu_{0})^{2} + \gamma^{2}} \right]
\end{equation}

\noindent where $f(\nu)$ is the rms-normalized spectral power density at frequency $\nu$, $A$ is the normalization of the power law component, $\alpha$ is the power law index, $B$ is the integrated power underneath the Lorentzian component, $\nu_0$ is the frequency at the peak of the Lorentzian component, and $\gamma$ gives the width of the Lorentzian and is related to the quality factor by $Q \equiv \frac{\nu_{0}}{2\gamma}$. We scanned $\nu_{0}$ over the frequency range, bounded on the lower end by the length of the intervals, and bounded on the upper end by the Nyquist frequency, equal to twice the binning frequency. For each value of $\nu_{0}$, we fit the average cospectrum to the model described by Equation \ref{eq:Lorentz_pl}, allowing all parameters aside from $\nu_{0}$ to vary, and compared the resulting $\chi^2$ fit statistic to the statistic attained by fitting the cospectrum to a power law alone. This produced a $\Delta\chi^2$ distribution over the range of allowed frequencies. The greatest fit improvement is $|\Delta\chi^2| < 12$, compared to a baseline statistic of $\chi_{0}^{2} = 2.5 \times 10^{5}$, indicating that the addition of a Lorentzian signal to the underlying power law spectrum is not warranted. We nonetheless attempted to fit QPO signals at several peaks in the $|\Delta\chi^2|$ distribution, this time allowing $\nu_{0}$ to vary. We found that none of these signals had significance greater than $1.7\sigma$, where significance is defined as the ratio of the integrated power under the Lorentzian to the error of the power, $B/\sigma_{B}$.

We performed a similar analysis for the burst emission as well as a more focused search for QPOs in two 10\,ks intervals preceding each burst. We produced cospectra for $115\times1$\,s intervals during Burst 1 and  $98\times1$\,s intervals during Burst 2, and averaged all 213 cospectra together. We note that for the purposes of timing analyses, we extended each of the burst GTIs by $75$\,s in order to include the burst decays. For the pre-burst emission, we produced a total of 34 cospectra, each corresponding to an interval of 256\,s, 18 of which precede Burst 1 and 16 of which precede Burst 2. We averaged over each of the pre-burst epochs separately, and, prompted by the work of \citet{Revnivtsev2001}, we filtered the pre-burst emission to include only photon energies in the range 3-12\,keV. Following the QPO search method we described for the persistent emission yields similar results: we do not significantly detect QPOs at any frequency during the bursts nor during the pre-burst intervals.

We also investigated the burst cospectra for evidence of coherent oscillations corresponding to a spin period. Visual inspection of both the dynamical and averaged (across both bursts) cospectra does not indicate the presence of coherent oscillations. We find a $99\%$ confidence upper limit of $16\%$ for the fractional rms integrated between $588$\,Hz and $590$\,Hz, exclusive.

We further searched for burst oscillations by performing a dynamical epoch-folding search on the burst emission. We analyzed each burst in 1 second intervals, stepping through the bursts in 0.5 second increments, resulting in overlapping time bins. For each focal plane module we produced pulse profiles corresponding to a range of oscillation frequencies near $589$\,Hz. We folded only events between $3$ and $12$\,keV, again restricting our analysis to events within the source extraction region used described above. The resulting pulse profiles were further corrected for variations in dead time using the method described in \citet{Madsen2015}. For each corrected pulse profile we calculated the $Z^{2}_{2}$ statistic \citep{Buccheri1983} as well as the corresponding probability that each of these statistics was not produced by noise. For each burst, we were thus left with two distributions, one for FPMA and another for FPMB, of the probability of detection of oscillations over a range of frequencies. We did not find any significant signal at any time during the bursts, and those spurious signals which we did observe only appeared in a single focal plane module.

\section{Discussion} \label{sec:discussion}

\comment{\subsection{Material composition and burst recurrence}
We have shown that Burst 2 as well as the second sub-burst of Burst 1 s}

\subsection{Explaining the double-peaked burst}

Both of the type-I X-ray bursts which we have presented exhibit interesting bursting behavior. Burst 1 does not adhere to the canonical fast-rise exponential-decay structure of most type-I bursts, instead showing a double-peaked structure. \replaced{Double-peaked bursts have been shown in some cases to be associated with PRE}{In some cases, bursts which show multiple peaks in their observed count rates have been shown to be associated with PRE} \citep{Lewin1984}, however we do not observe evidence for PRE during Burst 1\added{, as the bolometric luminosity exhibits the same double-peaked structure as the count rate}. Burst 1 is instead more similar to the multiple-peaked non-PRE bursts observed in sources like 4U 1608–52 \citep{Penninx1989,Jaisawal2019}\added{, GX 17+2 \citep{Kuulkers2002},} and 4U 1636-536 \citep{Sztajno1985,vanParadijs1986,Bhattacharyya2006}.

Even among this class of multiple-peaked bursters there is some inhomogeneity. The burst observed by Jaisawal et al. underwent rebrightening about 5\,s after burst onset, resulting in a double-peaked light curve where the second burst reached about half of the count rate as the first peak. The authors presented several possible physical explanations including a stalled burning front, waiting points in the rp-process, and reburning of material. \added{While the double-peaked burst which we have observed bears some dissimilarities with the double-peaked burst of 4U 1608–52,} these \added{mechanisms} could also explain the behavior of Burst 1. \deleted{However, the double-peaked burst which we have observed bears some dissimilarities with the double-peaked burst of 4U 1608–52. }

\deleted{First and foremost, }In the case of \grs, we have shown that the sub-bursts were characterized by quite different rise and decay times, and that the second sub-burst actually exceeded the peak luminosity of the first. Additionally, whereas Jaisawal et al. found that the blackbody temperature reached its peak during the dip in brightness between sub-bursts at the same time that blackbody radius reached a minimum, we find that the blackbody temperature closely tracks the luminosity, while appearing to show a slight anti-correlation with the blackbody radius. \deleted{The behavior that we have observed much more closely resembles a superposition of two separate bursts rather than rebrightening of a single burst.} This is reminiscent of the double- and triple-peaked bursts \replaced{observed by Sztajno et al. and van Paradijs et al}{originating from 4U 1636-536}. \deleted{Unfortunately in this case, the statistics were insufficient to fit the burst spectrum to such a superposition of two blackbody components.}

Similarly to \added{the scenario presented by} Sztajno et al. and van Paradijs et al., \deleted{we suggest a scenario wherein} the first sub-burst may \replaced{be}{have been} the result of a ``failed," slowly igniting type-I burst which did not manage to spread across the entirety of the NS surface, followed by\deleted{ (and perhaps triggering via heating of the surface)} a second type-I burst which ignited much more quickly and enveloped a larger fraction of the NS surface.\added{ Alternatively, the double-peaked structure we observed may be the result of a single burning front which stalled as it traveled from the NS pole towards the equator. Battacharyya \& Strohmayer have shown that models of burning front propagation are able to reproduce the qualitative features observed during a double-peaked burst in 4U 1636-536. Indeed, the shape and spectral evolution of the burst they present strongly resembles those of Burst 1. A third possibility is the stalling of burning due to waiting points in the process of thermonuclear burning \citep{Fisker2004}.} \deleted{A comparison of the integrated radiated energy of each burst demonstrates that the first sub-burst was somewhat less energetic than the second, consistent with a smaller amount of burned material: $E_{1,1}\approx2.3\times10^{38}\,\mathrm{erg}$ and $E_{1,2}\approx5.2\times10^{38}\,\mathrm{erg}$. For comparison, the energy radiated during Burst 2 was significantly greater than that of Burst 1 ($E_{1} = E_{1,1} + E_{1,2}\approx7.4\times10^{38}\,\mathrm{erg}$) at $E_{2}\approx1.6\times10^{39}\,\mathrm{erg}$.} \replaced{Further}{Detailed} physical modeling is required to \replaced{determine}{compare} the viability of \replaced{this}{these} scenario\added{s}, however this is beyond the scope of this paper.

\subsection{Photospheric Radius Expansion}

\begin{deluxetable*}{cccccc}
\tablenum{4}
\tablecaption{Comparison of PRE bursts presented in this work and in \cite{Barriere2015}.\label{tab:comparison}}
\tablewidth{0pt}
\tablehead{
\colhead{\multirow{2}{*}{Reference}} & \colhead{$F_{\mathrm{pers}}$\tablenotemark{a}} & \colhead{$F_{\mathrm{peak}}$\tablenotemark{b}} & \colhead{$X$\tablenotemark{c}} & \colhead{$L_\mathrm{Edd}$\tablenotemark{d}} & \colhead{$d$\tablenotemark{e}} \\
 & ($10^{-11}\,$\fluxcgs) & ($10^{-8}\,$\fluxcgs) & \nodata & ($10^{38}\,$\lumcgs) & (kpc)}
\startdata
\multirow{2}{*}{\cite{Barriere2015}} & \multirow{2}{*}{$8.04^{+0.38}_{-0.35}$} & \multirow{2}{*}{$3.58^{+0.29}_{-0.28}$} & $0$  & $3$   & $8.3\pm0.7$      \\
 &                                                                           &                                          & $0.7$ & $1.7$ & $6.3\pm0.5$      \\
\noalign{\smallskip}
\multirow{2}{*}{This work} & \multirow{2}{*}{$130 \pm 2$} & \multirow{2}{*}{$2.94^{+0.28}_{-0.26}$} & $0$   & $3$   & $9.0\pm0.5$      \\
 & &                                                                                                & $0.7$ & $1.7$ & $7.0 \pm 0.4$      \\
\noalign{\smallskip}
\enddata
\tablenotetext{a}{\footnotesize Unabsorbed bolometric flux calculated for the period leading up to the burst. This period corresponds to epoch O4 in \cite{Barriere2015} and to the final four orbits of the observation presented in this work.}
\tablenotetext{b}{\footnotesize Unabsorbed bolometric flux calculated for  time slice corresponding to the peak of the burst.}
\tablenotetext{c}{\footnotesize Hydrogen mass fraction assumed when calculating the Eddington luminosity.}
\tablenotetext{d}{\footnotesize Eddington luminosity calculated using Equation \ref{eq:Edd_Lewin}}
\tablenotetext{e}{\footnotesize Distance calculated assuming that the peak flux corresponds to the Eddington luminosity.}
\end{deluxetable*}

Our analysis of the spectral evolution of Burst 2 revealed that the burning material underwent a process of rapid expansion and contraction. We have interpreted this behavior as evidence for PRE, during which the burning material reached the Eddington luminosity and was therefore momentarily lifted from the NS surface by radiation pressure. The evolution of the blackbody radius, temperature, and luminosity strongly resembles the PRE burst observed by \citet{Barriere2015}, with the major difference being the overall shorter timescales observed in Burst 2 compared to the Hydrogen-rich burst observed by those authors.

Compared to the burst observed by Barrière et al., the short rise and decay time observed for Burst 2, in addition to the relatively high persistent flux leading up to the burst (see Table \ref{tab:comparison}) may indicate that at the time of observation the source was in the stable H-burning regime\replaced{, leading}{. In other words, while the accreted material is likely to contain a mixture of H and He, the accreted H burns stably on the NS surface rather than leading to runaway thermonuclear burning. This in turn leads} to pure He bursts \citep{Trap2009}. \added{In addition to the burst structure and persistent flux, the recurrence timescale and burst fluence provide further evidence for pure He burning during Burst 2. The recurrence timescale, $\tau_\mathrm{rec}$, and the mass accretion rate, $\dot{m}$, can be combined to calculate the ignition column: $y_\mathrm{ign}=\dot{m}\tau_\mathrm{rec}(1+z)^{-1}$. Next, the energy released per unit mass, $\epsilon_\mathrm{nuc}$, can be calculated using the relation $\epsilon_\mathrm{nuc}=E_\mathrm{b}(1+z)/(4\pi R_\mathrm{NS}^{2}y_\mathrm{ign})$, where $E_\mathrm{b}$ is the total energy radiated during the burst and is related to the burst fluence, $f_\mathrm{b}$, by $E_\mathrm{b}=4\pi d^2 f_\mathrm{b}$. We may combine the relations listed above with Equation \ref{eq:m_dot} to obtain an equation for $\epsilon_\mathrm{nuc}$ which does not depend on the distance to the source but rather on the ratio of the observed burst fluence to the integrated persistent flux:}

% \begin{equation}
% \begin{aligned}
%     \epsilon_{\mathrm{nuc}}= & 2.44 \times \left(\frac{f_{\mathrm{b}}}{10^{-7}\,\mathrm{erg\,cm^{-2}}}\right)  \left(\frac{\tau_{\mathrm{rec}}}{10^{4}\,\mathrm{s}}\right)^{-1}\\ 
%     & \times \left(\frac{F_{\mathrm{bol}}}{10^{-9}\,\mathrm{erg\,cm^{-2}\,s^{-1}}}\right)^{-1} \left(\frac{M}{1.4\,M_{\odot}}\right)\\
%     & \times \left(\frac{R_{\mathrm{NS}}}{10\,\mathrm{km}}\right)^{-1} \left(\frac{1+z}{1.31}\right) \times 10^{18}\,\mathrm{erg\,g^{-1}}
% \end{aligned}
% \label{eq:eps_nuc}
% \end{equation}

\begin{equation}
\begin{aligned}
    \epsilon_{\mathrm{nuc}}= & 2.44 \times \left(\frac{f_{\mathrm{b}}}{10^{-7}\,\mathrm{erg\,cm^{-2}}}\right)  \left(\frac{\tau_{\mathrm{rec}}}{10^{4}\,\mathrm{s}}\right)^{-1}\\ 
    & \times \left(\frac{F_{\mathrm{bol}}}{10^{-9}\,\mathrm{erg\,cm^{-2}\,s^{-1}}}\right)^{-1} \times 10^{18}\,\mathrm{erg\,g^{-1}}
\end{aligned}
\label{eq:eps_nuc}
\end{equation}

\noindent \added{for a NS with $M=1.4M_{\odot}$ and $R_\mathrm{NS}=10\,\mathrm{km}$. Due to the orbital gaps in the observation, we were only able to determine an upper limit on the recurrence timescale, $\tau_\mathrm{rec} \lesssim 4\times10^4\,\mathrm{s}$. Integrating the blackbody flux during Burst 2 gives a burst fluence of $f_\mathrm{b}=(2.8 \pm 0.1)\times 10^{-7}\,\mathrm{erg\,cm^{-2}}$. Thus we arrive at an energy per unit mass of $\epsilon_\mathrm{nuc} \gtrsim 1.4\times10^{18}\,\mathrm{erg\,g^{-1}}$ or an energy per nucleon of $Q_\mathrm{nuc}\gtrsim1.4\,\mathrm{MeV\,nucleon^{-1}}$. Given $Q_\mathrm{nuc}=(1.35+6.05X)\,\mathrm{MeV\,nucleon^{-1}}$ \citep{Goodwin2019}, this corresponds to a H fraction, $X$, consistent with zero.}

With this in mind, we may calculate the distance to the source. The Eddington luminosity as measured by a distant observer is given by \citep{Lewin1993}

\begin{equation}
\begin{aligned}
    L_{\mathrm{Edd},\infty} = & \frac{4 \pi cGM}{\kappa_{0}} \left[ 1-\frac{2GM}{Rc^{2}} \right]^{1/2} \\
        & \times \left[ 1+ \left( \frac{kT}{39.2\,\mathrm{keV}} \right)^{0.86} \right] (1+X)^{-1}
\end{aligned}
\label{eq:Edd_Lewin}
\end{equation}

\noindent where $\kappa_{0}$ is the opacity for pure He and is given by $\kappa_{0}=0.2\ \mathrm{cm^{2}\,g^{-1}}$, $M$ is the mass of the accretor, $R$ is the radius at which the Eddington luminosity is being calculated, $T$ is the temperature of the material at touchdown, and $X$ is the hydrogen mass fraction of the material, which we assume to be zero. We assume a typical NS mass of $M=1.4M_{\odot}$, and because the peak flux is achieved at the time of touchdown when the burning material is assumed to have returned to the NS surface, we similarly assume a typical NS radius of $R=10\,\mathrm{km}$. The temperature, $T$, that appears in Equation \ref{eq:Edd_Lewin} is that of the material in the emission reference frame. In reality, the temperature measured by a distant observer via fitting to a blackbody \replaced{is cooler than}{differs from} the temperature in the frame of the emitting material, $T_\mathrm{emit}=(1+z)f_\mathrm{c}^{-1}T_{\mathrm{\infty}}$. Given a temperature at touchdown of $kT_{\mathrm{\infty}}=2.44$\,keV, a redshift of $(1+z)=1.31$ (see Section \ref{subsec:persistent_spectra}), \added{and a color correction factor of $f_\mathrm{c}=1.7$,} we arrive at an Eddington luminosity of $L_{\mathrm{Edd},\infty}=3\times10^{38}\,\mathrm{erg\,s^{-1}}$. Combined with our measured peak unabsorbed, bolometric flux of $F_\mathrm{peak}=2.94^{+0.28}_{-0.26}\times10^{-8}\,\mathrm{erg\,cm^{-2}\,s^{-1}}$, this corresponds to a distance to the source of $d=9.0\pm0.5$\,kpc.

This result is consistent with previous estimates which have placed the source at distances ranging from $5$\,kpc to $9$\,kpc depending upon the assumed material composition and the method of calculating the Eddington luminosity \citep{Cocchi1999,Galloway2008,Trap2009,Barriere2015}. Our measurement lies on the most distant end of these estimates. This could be the result of a number of effects. As we have shown, the material composition has a significant effect on the Eddington luminosity and therefore on any measurement of the distance. For example, if we assume solar composition, $X=0.7$, rather than pure He burning, the resulting distance is $d=7.0 \pm 0.4$\,kpc. \added{In order to remain consistent with our measurements of the burst fluence, this would require a recurrence time which is a factor of $\sim4$ smaller than the time which elapsed between Burst 1 and Burst 2. This in turn would imply that an additional burst took place during one of the orbital gaps of the observation.} It is interesting to note that the peak flux we have observed is nearly consistent with the peak flux presented by Barriere et al., while the structures of the corresponding bursts are very different. This may represent a challenge to the method of inferring the composition of burning material via light curve analysis alone. 

\replaced{Additionally, the peak flux of type-I bursts can vary due to changing disk inclination \citep{Galloway2003}.}{Throughout this paper we have assumed isotropic emission when calculating burst and persistent fluxes. However, the inclination of the system can have a significant effect on the actual values of these quantities \citep{Galloway2003}. Spectral modeling of the persistent emission suggests that the inclination of the inner disk is less than $30^{\circ}$. Depending upon the disk geometry, this may result in an overestimate of the burst flux by a factor of $1.5-2$, and of the persistent flux by up to a factor of $2.5$ \citep{He2016}. Therefore, the distance may be underestimated, both in this work and in previous work which also assumed isotropic emission, by a factor of up to $\sqrt{2}$.} \deleted{Indeed,}The bursting behavior presented by \citet{Trap2009} indicates a possible correlation between peak flux and burst fluence\deleted{(the total energy radiated during a burst)}. We therefore speculate that those bursts could actually have achieved the same peak luminosity (modulo composition effects) which only appeared to vary due to changes in inclination. However, because Trap et al. are unable to claim that the bursts undergo PRE, it may not be appropriate to assume that the bursts reached the Eddington luminosity, meaning that this interpretation may not be accurate. Future investigations of the long-term variability of \grs\ could reveal whether changing disk inclination has a significant impact on the measured peak flux of PRE bursts.

\section{Summary and Conclusions} \label{sec:conclusions}

We have presented light curve, spectral, and timing analyses of two type-I X-ray bursts as well as persistent emission originating from the source \grs\ during a period of outburst observed by \nustar. \added{This represents the first time the source has been observed by \nustar\ at such a high persistent flux -- the source was over 15 times brighter than during the previous \nustar\ observation which took place in 2013 -- which allowed us to analyze the bursting behavior in an accretion regime previously unexplored by \nustar\ for this source. Additionally, because we observed two bursts, we were able to obtain an upper limit on the recurrence timescale between the bursts, $\tau_\mathrm{rec} \lesssim 4\times10^4\,\mathrm{s}$. This upper limit is half of the shortest recurrence time previously reported between bursts originating from \grs\ \citep{Trap2009}, and the measurement of the recurrence time served as an additional piece of evidence that the burning material was H-deficient.} Aside from the two bursts, the source appeared to show only slight variability. We confirmed this by analyzing the cospectra taken between light curves observed by the two focal plane modules. In all cases we found the cospectra to be relatively featureless, being described well by a single power law model. Additionally, we performed an epoch-folding analysis which did not produce evidence for coherent burst oscillations.

We found that the spectra of the persistent emission were modeled well by an absorbed blackbody plus upscattered emission originating from a hot corona or accretion flow, in addition to Fe K$\alpha$ emission which can be modeled either phenomenologically using a simple Gaussian emission line or self-consistently using relativistic disk reflection models. Having modeled the persistent emission, we were able to perform a time-resolved spectroscopic analysis on each of the type-I bursts. We found that the first burst exhibited a double-peaked structure, and we showed that the structure we observed did not appear to result from the rebrightening of an ongoing burst but rather could be described as two ``sub-bursts" occurring in quick succession. \added{This is the first confirmed case of such a burst originating from \grs, placing it within a population of only a handful of multiple-peaked bursters.}

The first burst was significantly less bright and energetic than the second burst, which exhibited a canonical fast-rise exponential decay structure in addition to an extended tail. Spectroscopic analysis showed that the blackbody describing the emission during the second burst underwent a period of rapid contraction accompanied by a rise in temperature, all occurring at a constant luminosity. We therefore inferred that the source had undergone photospheric radius expansion, or PRE, which typically implies that the source has reached the Eddington luminosity. Based on this assumption, we calculated a distance to the source of $d=9.0\pm0.5$\,kpc, which is somewhat larger than, but still consistent with, previous estimates. However, the value of the Eddington luminosity depends strongly on the composition of the burning material\replaced{, which we are only able to infer based on the persistent flux and the burst shape.}{. While the observed properties of the burst as well as the persistent flux and recurrence timescale are consistent with pure He burning, it is not possible to determine the composition with absolute certainty.} Future observations type-I X-ray bursts originating from \grs\ may provide additional opportunities to determine how the composition of burning material varies with persistent flux and how this affects bursting behavior, as well as whether changing disk inclination is responsible for variation in peak flux between PRE bursts at different times.

\acknowledgments
Jiachen Jiang acknowledges support by the Tsinghua Astrophysics Outstanding (TAO) Fellowship and the Tsinghua Shuimu Scholar Programme. R. M. Ludlam acknowledges the support of NASA through Hubble Fellowship Program grant HST-HF2-51440.001. D. J. K. Buisson acknowledges support from the Royal Society. J. A. Garc\'ia acknowledges support from NASA ATP grant 80NSSC20K0540, and from the Alexander von Humboldt Foundation. This work was partially supported under NASA contract No. NNG08FD60C and made use of data from the \nustar\ mission, a project led by the California Institute of Technology, managed by the Jet Propulsion Laboratory, and funded by the National Aeronautics and Space Administration. We thank the \nustar\ Operations, Software, and Calibration teams for support with the execution and analysis of these observations. This research has made use of the \nustar\ Data Analysis Software (NuSTARDAS), jointly developed by the ASI Science Data Center (ASDC, Italy) and the California Institute of Technology (USA). \added{We would also like to acknowledge the insightful feedback provided by the reviewer, which significantly improved the quality of this work.}
\software{Astropy, DS9, Scipy, Stingray, XSpec}

\bibliography{main}{}
\bibliographystyle{aasjournal}


\listofchanges

\end{document}